\documentclass[prb,twocolumn,showpacs]{revtex4}

\usepackage{graphicx,epsf}
\usepackage{bm, bbm}      
\usepackage{amsmath}
\usepackage{amssymb}
\usepackage{braket}
\usepackage{multirow}
\usepackage{relsize}

\usepackage{float}

\usepackage{comment}

\newcommand{\bone}{\mathbbm{1}}
\newcommand{\mos}{MoS$_2$}

\newcommand{\Hhat}{\hat{H}}
\DeclareMathAlphabet{\mathitbf}{OML}{zplm}{b}{it}
\renewcommand\vector[1]{\mathbf{#1}}

\newcommand\Av{\vector{A}}

\newcommand\Bv{\vector{B}}

\newcommand\kv{\vector{k}}
\newcommand\Kv{\vector{K}}

\newcommand\qv{\vector{q}}

\newcommand{\atilde}{\tilde{a}}

\newcommand{\psit}{\tilde{\psi}}
\newcommand\er{\,\mathrm{e}}

\newcommand{\eps}{\varepsilon}

\newcommand{\beq}{\begin{equation}}
\newcommand{\beqn}{\begin{eqnarray}}
\newcommand{\eeq}{\end{equation}}
\newcommand{\eeqn}{\end{eqnarray}}
\newcommand{\nn}{\nonumber}

\newcommand{\MoS}{MoS$_2$}

\usepackage{color}

\DeclareMathOperator{\h}{h}

\begin{document}

\def\tende#1{\,\vtop{\ialign{##\crcr\rightarrowfill\crcr
\noalign{\kern-1pt\nointerlineskip}
\hskip3.pt${\scriptstyle #1}$\hskip3.pt\crcr}}\,}

\title{Spin- and valley-dependent magneto-optical properties of \mos}
\author{F\'{e}lix Rose$^{1,2}$, M. O. Goerbig$^{1}$ and Fr\'ed\'eric Pi\'echon$^{1}$}

\affiliation{
$^1$Laboratoire de Physique des Solides, CNRS UMR 8502, Univ. Paris-Sud, F-91405 Orsay cedex, France\\
${^2}$Physics Department, \'Ecole polytechnique, 91128 Palaiseau, France}

\begin{abstract}
We investigate the behavior of low-energy electrons in two-dimensional molybdenum disulfide
when submitted to an external magnetic field. Highly degenerate Landau levels form in the material, between which light-induced excitations are possible. The dependence of excitations on light polarization and energy is explicitly determined, and it is shown that it is possible to induce valley and spin polarization, i.e. to excite electrons of selected valley and spin. Whereas the effective low-energy model in terms of massive Dirac fermions yields dipole-type selection rules, higher-order band corrections allow for the observation of additional transitions.
Furthermore, inter-Landau-level transitions involving the $n=0$ levels provide a reliable method for an experimental measurement of the gap and the spin-orbit gap of molybdenum disulfide.
\end{abstract}
\maketitle

\section{Introduction}

Molybdenum disulfide (\mos), in its two-dimen-sional (2D) form, has recently been isolated via the 
exfoliation technique, similarly to graphene and 2D boron nitride.\cite{novoselovPnas} 
In contrast to bulk or few-layer \mos, which is an indirect-gap semiconductor, 
recent experiments have shown that 2D \mos{} is a semiconductor with a direct gap on the order 
of 1.66~eV,\cite{makPrl,exp1} in agreement with ab-initio and tight-binding 
calculations.\cite{mattheiss,lebPrb,yunPrb,cheiPrb,rosArx,roldanArx,ochArx,korArx}
The direct gap is situated at the corners $\Kv$ and $\Kv'=-\Kv$ of the hexagon-shaped first
Brillouin zone. In the vicinity of these points (valleys)
and at low energy, the electronic properties can be modelled by massive Dirac fermions with a moderate spin-orbit 
coupling.\cite{xiaoPrl,roldanArx} This opens the fascinating possibility to study the particular
topological properties of pseudo-relativistic fermions in another condensed-matter system
than graphene where the low-energy electronic properties are governed by massless Dirac
fermions.\cite{antonioRev} Recent experiments have indeed shown\cite{zengNN,makNN} that circularly polarized
light allows one to address electrons in a single valley, in agreement with previous analytical\cite{xiaoPrl}
and numerical ab-initio\cite{caoNC} calculations, such that \mos~might be a promising candidate for
valleytronics devices.

One of the most salient features of 2D Dirac fermions in condensed-matter systems is certainly
their topological property in the form of a singularity in the wave function at the (massive)
Dirac point that gives rise to a non-zero Berry curvature.\cite{niuPrl,fuchsEpjb} A prominent
consequence is an anomaly in the $n=0$ Landau level in the presence of a magnetic field, which in
contrast to all other levels is bound either to the top of the valence or the bottom of the 
conduction band. 

In the present paper, we study the magneto-optical properties of 2D \mos, which arise precisely
due to the presence of massive Dirac fermions in the vicinity of the $\Kv$ points. Whereas the 
$n=0$ Landau level at the $\Kv$ point is situated at the top of the valence band, that at the 
$\Kv'$ point is bound to the bottom of the conduction band. As a consequence, circularly 
polarized light allows one to excite electrons in a single valley if the inter-Landau-level transition
involves the $n=0$ level.  This transition provides a direct measure of the mass gap and the 
spin-orbit coupling in \mos.  Very similar results have recently been obtained by Tabert and 
Nicol, who investigated the magneto-optical properties of silicene and similar 2D materials
that may also be described in terms of massive Dirac fermions at low energy.\cite{tabArx,tabArx2}
Whereas massive Dirac fermions respect the dipole selection rules 
$n\rightarrow n\pm 1$ (regardless of the band), we show furthermore 
that higher-order band corrections such 
as trigonal warping give rise to novel allowed transitions, such as the inter-band transitions
$n\rightarrow n$ or $n\rightarrow n\pm 2$ and $n\rightarrow n\pm 4$.

The remaining parts of the paper are organized as follows. In Sec. \ref{sec:GO}, we build up the 
model Hamiltonian, which is discussed in the absence and the presence of a magnetic field, the latter 
giving rise to the Landau-level spectrum. The magneto-optical excitations within the model of
massive Dirac fermions with spin-orbit coupling are investigated in Sec. \ref{sec:pol}, where
the selection rules for \mos~are presented. Section \ref{sec:TW} is devoted to the study of 
band corrections, their influence on the Landau-level spectrum, and novel optical transitions beyond
the dipole-allowed ones.

\begin{figure}[htb]
\centering
\includegraphics[width=7cm,angle=0]{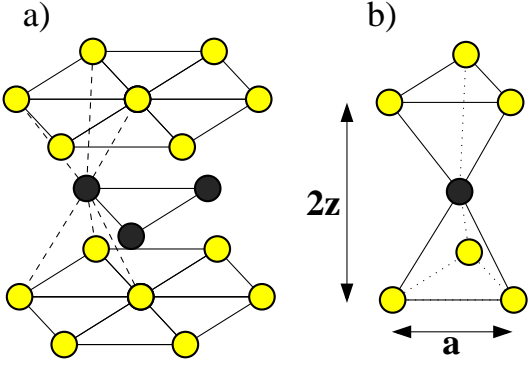}
\caption{\footnotesize{ Crystal structure of 2D $\mathrm{MoS}_2$. a) Perspective view of the honeycomb lattice.  b) The unit cell. Each of the two sulphur atoms of the A sub-lattice (purple) are separated by a length\cite{yunPrb} $z\approx 1.572$~\AA~from the B sub-lattice (red) plane. The characteristic lattice spacing\cite{yunPrb} $\tilde{a}\approx 3.16$~\AA~is defined as the distance between the two closest points in a sub-lattice.
}}
\label{fig:lay}
\end{figure}

\section{General overview}
\label{sec:GO}
In this section, we characterize both the low-energy behaviour of electrons in $\mathrm{MoS_2}$ in the absence and the presence of a transverse magnetic field. We begin by introducing the Dirac Hamiltonian and the spin-orbit coupling, before considering the effect of a magnetic field.

\subsection{Low-energy Hamiltonian}

The electronic behavior of $\mathrm{MoS_2}$ has been studied both in the framework
of ab-initio and tight-binding calculations.\cite{rosArx,roldanArx,korArx,xiaoPrl,lebPrb} The material has one molybdenum and two sulphur atoms per unit cell (see Fig. \ref{fig:lay})
and in total eleven orbitals thus need to be considered: three $p$ orbitals for each of the sulphur and five $d$ orbitals from the molybdenum atoms. In contrast to bulk or few-layer
$\mathrm{MoS_2}$, which are indirect-gap semiconductors,\cite{mattheiss,yunPrb,cheiPrb} a single layer of $\mathrm{MoS_2}$ has 
a direct gap of roughly 1.66~eV at the $\Kv$ and $\Kv '=-\Kv$ points situated at the corners of the 
hexagonal first Brillouin zone.\cite{yunPrb} In spite of the complexity of the band structure, the
low-energy electronic properties of \MoS, in the vicinity of the two valleys $\Kv$ and $\Kv '$,
may be understood within a simplified model that only takes into
account three molybdenum orbitals: $\ket{d_\mathrm{3r^2-z^2}}$, which mostly forms the bottom of the 
conduction band, and a valley-dependent mix of $\ket{d_\mathrm{xy}}$ and $\ket{d_\mathrm{ x^2- y^2}}$
for the top of the valence band,\cite{xiaoPrl}

\beq
\ket{\phi_c}=\ket{d_\mathrm{3r^2-z^2}} \text{, }
\ket{\phi_v^\xi}=\frac{1}{\sqrt{2}}\left(\ket{d_\mathrm{x^2-y^2}}+i\xi\ket{d_\mathrm{xy}}\right)\text{.}
\eeq
Here, $\xi=+$ denotes the valley $\Kv$ and $\xi=-$ stands for $\Kv'$.
If one represents the Hamiltonian in this basis and expands it around the points 
$\Kv$ and $\Kv '$, the low-energy Hamiltonian of the system can be written as\cite{xiaoPrl}

\beq
\hat{H}_0=\hbar v_F \left(\xi q_x \sigma_x + q_y \sigma_y \right) +\Delta \sigma_z\text{,}
\eeq
in which $\sigma_x, \sigma_y$, and $\sigma_z$ are Pauli matrices, 
$\qv$ is the reciprocal lattice vector measured with respect to $\xi \Kv$ with 
$\atilde\left|\qv\right| \ll 1$ ($\atilde$ being the characteristic lattice spacing). Notice that 
the Fermi velocity $v_F\approx 85\,000$~m/s in $\mathrm{MoS_2}$ is comparable to that of graphene.

\subsection{Spin-orbit coupling}
\begin{figure}[t]
\centering
\includegraphics[width=5.5cm,angle=0]{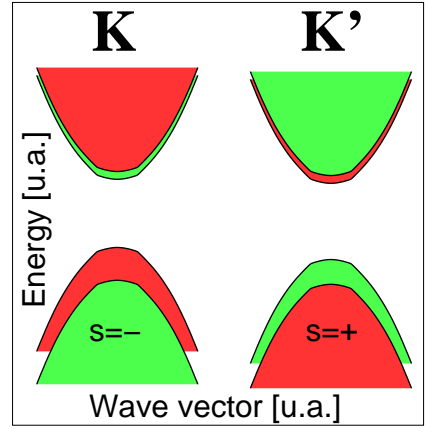}
\caption{\footnotesize{ (Color online) Sketch of the low-energy band structure of molybdenum disulfide in arbitrary units (u.a.). The band structure consists in two pairs of massive Dirac cones separated by spin-orbit coupling.}}
\label{fig:bnd}
\end{figure}

As mentioned above, $\mathrm{MoS_2} $ is characterized by a strong intrinsic spin-orbit coupling
(see Fig. \ref{fig:bnd}). The spin-orbit Hamiltonian, which needs to be added to $\Hhat_0$, is

\beq
\Hhat_{\mathrm{so}}=\xi 
\begin{pmatrix}
\Delta_\mathrm{so}^c & 0\\
0 & \Delta_\mathrm{so}^v
\end{pmatrix}
\otimes s_z,
\eeq
in which $s_z$ is the Pauli matrix for spin (of eigenvalues $\pm1$) and $2\Delta_\mathrm{so}^{c,v}$ 
is the spin-orbit gap in the conduction and valence band, respectively. 
Ab-initio calculations indicate that 
$\Delta_\mathrm{so}^v \approx 150$~meV while $\Delta_\mathrm{so}^c$ is very small but finite 
($\Delta_\mathrm{so}^c \approx 3$~meV). Thus, we note 
\beq
\Delta_\mathrm{so}=\Delta_\mathrm{so}^v-\Delta_\mathrm{so}^c \approx \Delta_\mathrm{so}^v.
\eeq
Obviously, as long as spin relaxation processes are not considered, the spin remains a good 
quantum number. It is noteworthy that $\Delta_\mathrm{so}^{c,v} \ll \Delta $, and thus the low-energy 
physical properties in $\mathrm{MoS_2}$ are largely controlled by the mass gap $2\Delta$. 
Therefore, in spite of the similarity with the model Hamiltonian used in the description of graphene 
with a spin-orbit gap\cite{KaneMele} or silicene,\cite{ezawaPrl}
no quantum spin Hall effect is to be expected in 
$\mathrm{MoS_2}$ because the latter would require $\Delta_\mathrm{so}^{c,v} \gg \Delta $. 
Even if $\Hhat_\mathrm{so}$ is different in each valley, it is locally constant around $\xi \Kv $ 
and therefore does not complicate the analysis of the orbital (wave-vector dependent) electronic properties, such as the calculation of the Landau levels (see Sec. \ref{sec:LL}). Thus, the system is equivalent to two spin-resolved Dirac Hamiltonians similar to $\hat{H}_0$ with a spin- and valley-dependent gap $2\Delta_{\xi s}$ as well as a constant energy term, 

\beqn
\Hhat_{\xi, s}&=&\Hhat_0(\Delta=\Delta_{\xi s})+\frac{\Delta_\mathrm{so}^c+\Delta_\mathrm{so}^v}{2} \bone_2 \label{eq:hso} \\
\Delta_{\xi s}&=&\Delta-\frac{\xi s}{2} \Delta_\mathrm{so} \label{eq:delt}.
\eeqn
The term of constant energy $(\Delta_\mathrm{so}^c+\Delta_\mathrm{so}^v)\bone_2/2$ plays no
physical role and is omitted henceforth.

\subsection{Landau levels}
\label{sec:LL}

When 2D electrons are subjected to a transverse magnetic field $\Bv=\nabla\times\Av$, Landau levels form and the energies within the valence and conduction bands get quantized. Indeed, making the Landau-Peierls substitution $\qv \rightarrow \qv+e\Av/\hbar$ in the Hamiltonian shows that it is possible to write the wave functions of the Hamiltonian as $\ket{\psi_\kv}=\phi(x)\exp(ik_yy)$ in which $\phi$ is an eigenvector of the effective Hamiltonian
\beq
\Hhat_\Bv^{\xi,s}= \hbar v_F \left[\xi q_x \sigma_x +\left(k_y+\frac{eB}{\hbar}{x}\right)\sigma_y\right]+ \Delta_{\xi s} \sigma_z.
\eeq
where we have used the Landau gauge $\Av=(0,Bx,0)$ for the vector potential. Because $x$ and $q_x$ do not commute, it is possible to rewrite $\Hhat_\mathrm{eff}$ using dimensionless operators $Q=l_B q_x$ and $X=x/l_B+\hbar k_y/eB$ such that $\left[X,Q \right]=i$. Here, $l_B=\sqrt{\hbar/eB}$ is the magnetic length.

\beq
\Hhat_\Bv^{\xi,s}= \frac{\hbar v_F}{l_B} (\xi Q\sigma_x +X \sigma_y )+\Delta \sigma_z
\eeq
With the help of the ladder operators $a=(X+iQ)/\sqrt{2}$ and $a^\dagger=(X-iQ)/\sqrt{2}$ we may rewrite $\Hhat_\mathrm{eff}$ in both valleys,

\beqn
\Hhat_\Bv^{\xi=+,s} &=&
\begin{pmatrix}
\Delta_s & -i \eps a \\ 
i \eps a^\dagger & -\Delta_s
\end{pmatrix},
\\
\Hhat_\Bv^{\xi=-,s}&=&
\begin{pmatrix}
\Delta_{-s} & i \eps a^\dagger \\ 
-i \eps a & -\Delta_{-s}
\end{pmatrix},
\eeqn
in which $\eps = \sqrt{2} \hbar v_F/l_B\approx 30.5 \, \sqrt{B(T)}$~meV.

Using the eigenvectors $\ket{n}$ of the number operator $n=a^\dagger a$ it is possible to find the eigenstates of the Hamiltonian in both valleys
\beqn
\psi_{\lambda n}^{\xi=+,s}&=&(\alpha_{\lambda n}^{+ s} \ket{ n-1 } , \beta_n \ket{n} ) \text{ for } n \geqslant 1, \label{vectors1}\\
\psi_{-0}^{\xi=+,s}&=& (0,\ket{0}) \text{ for } n=0,\\
\psi_{\lambda n}^{\xi=-,s}&=&(\alpha_{\lambda n}^{- s} \ket{ n } , \beta_n \ket{n-1} ) \text{ for } n \geqslant 1,\\
\psi_{+0}^{\xi=-,s}&=& (\ket{0},0) \text{ for } n=0. \label{vectors2}
\eeqn
where $\lambda=\pm 1 $ designates the band. Here, the coefficients $\alpha_{\lambda n}^{\xi s}$ and $\beta_n$ are defined as

\beqn
\alpha_{\lambda n}^{\xi s} &=& \Delta_{\xi s}+\lambda\sqrt{\Delta_{\xi s}^2+n\eps^2},\\
\beta_n &=& -i\sqrt{n}\eps .
\eeqn
Counting possibles values of $k_y$ yields that the Landau-level degeneracy 
is $n_B=eB/h$ for each of the four spin-valley branches.

Notice that the norm of the vector $\psi_{\lambda n}^{\xi,s}$ is the same for both valleys and will be noted as $N_{\lambda n}^{\xi s}$
\beqn\label{eq:normal}
\nn
N_{\lambda n}^{\xi s}&=&\sqrt{\left|\alpha_{\lambda n}^{\xi s}\right|^2+\left|\beta_n\right|^2}\text{ for }n \geqslant 1,\\
N_{\lambda 0}^{\xi s}&=&1.
\eeqn
The energy associated with the spinor $\psi_{\lambda n}^{\xi,s}$ is 

\beqn
\eps_{\lambda n}^{\xi s}&=& \lambda\sqrt{\Delta_{\xi s}^2+n\eps^2} \text{ for } n \geqslant 1.
\eeqn

\begin{figure}[t]
\centering
\includegraphics[width=5.5cm,angle=0]{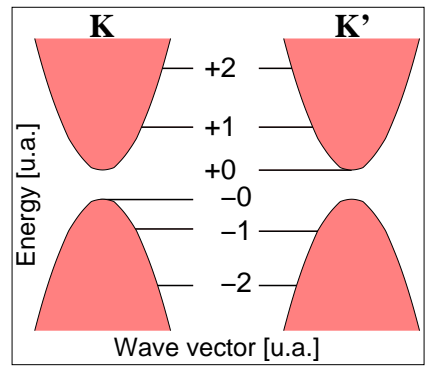}
\caption{\footnotesize{(Color online) Representation of the Landau levels in the valleys $\Kv$ and $\Kv' $ when spin-orbit coupling is not taken into account, i.e. it is assumed that $\Delta_{\xi s}=\Delta$. In that peculiar case, the energy levels are spin-degenerated. For the sake of clarity, the figure is not to scale as the energy separating the closest Landau levels is way smaller than the gap.}}
\label{fig:LL}
\end{figure}

In contrast to the $n \neq 0$ levels, which occur in pairs in each valley (one for each band), the $n=0$ level needs to be treated apart. Indeed, one finds a single $n=0$ level per valley. In the present case, as $\Delta_{\xi s}$ is always positive (since $\Delta_\mathrm{so} \ll \Delta $), for both values of spin the $n=0$ Landau levels in the $\Kv$ valley are fixed at the top of the valence band ($\eps_{n=0}^{+s}=-\Delta_{+ s}$) whereas the two $n=0$ levels in the $\Kv '$ valley are located at the bottom of the conduction band ($\eps_{n=0}^{-s}= \Delta_{- s}$) [see Fig. \ref{fig:LL}]. This is a direct consequence of the fact that electrons behave as massive Dirac fermions. The two valleys react differently to the magnetic field, and the particular behavior of the $n=0$ Landau levels is due
to the particular winding properties of the Berry phase, as may be understood in the framework of 
a semiclassical analysis.\cite{fuchsEpjb}

Notice finally that, if $2\Delta_\mathrm{so} > \Delta$, $\Delta_{+}$ is negative while $\Delta_{-}$ remains positive. Therefore, in both valleys, the $\psi^{\xi,s=+}_{\pm 0}$ states would be fixed to the bottom of the conduction band and the $\psi^{\xi,s=-}_{\pm 0}$ states are at the top of the valence band, which is a case discussed in the framework of silicene.\cite{tabArx,tabArx2}

\begin{table*}[t]
{
\centering
\begin{tabular}{|c||c|c|c|c|}
\hline
\multirow{2}{*}{Transitions} & \multicolumn{4}{c|}{Valley and light polarization}\\
\cline{2-5}
 & $\xi=+1 \text{, } \sigma=+1$ &  $\xi=+1 \text{, } \sigma=-1$ &$\xi=-1 \text{, } \sigma=+1$ & $\xi=-1 \text{, } \sigma=-1$
\\ \hline \hline
$\psi^{\xi,s}_{-(n+1)}\rightarrow \psi^{\xi,s}_{\lambda n}$ & $0$ &  $\left|\frac{\alpha_{-(n+1)}^{+s}\beta_n}{N_{-(n+1)}^{+s}N_{\lambda n}^{+s}}\right|^2$ &0&$\left|\frac{\alpha_{\lambda n}^{- s} \beta_{n+1}}{N_{-(n+1)}^{-s}N_{\lambda n}^{-s}}\right|^2$ \\  \hline

$\psi^{\xi,s}_{\lambda n}\rightarrow \psi^{\xi,s}_{n+1}$ & $\left|\frac{\alpha_{ n+1}^{+s}\beta_n}{N_{\lambda n}^{+s}N_{n+1}^{+s}}\right|^2$ &  $0$  & $\left|\frac{\alpha_{\lambda n}^{- s} \beta_{n+1}}{N_{\lambda n}^{-s}N_{n+1}^{-s}}\right|^2$&0\\ \hline
$\psi^{\xi,s}_{-1}\rightarrow \psi^{\xi,s}_{0}$ & $0$ &  $\left|\frac{\alpha_{-1}^{+s}}{N_{-1}^{+s}}\right|^2$ &0&$\left|\frac{\beta_1}{N_{-1}^{-s}}\right|^2$ \\ \hline
$\psi^{\xi,s}_{0}\rightarrow \psi^{\xi,s}_{+1}$ & $\left|\frac{\alpha_{+1}^{+s}}{N_{+1}^{+s}}\right|^2$ &  0  & $\left|\frac{\beta_1}{N_{+1}^{-s}}\right|^2$&0 \\ \hline
\end{tabular}}
\caption{\footnotesize Values of relative transition rates $\mathcal{P}_{\lambda_i n_i,\lambda n_f}^{\xi, s, \sigma}$ for every possible transition $\ket{i}\rightarrow\ket{f}$ are given as a function of the valley and light polarization. Here, $n$ denotes a non-zero positive integer and the state labelled as $\psi^{\xi,s}_{0}$ is either $\psi^{\xi=+1,s}_{-0}$ in the $\Kv$ valley or $\psi^{\xi=-1,s}_{+0}$ in the $\Kv '$ valley.
\label{tab:trans}}
\end{table*}

\section{Magneto-optical excitations}
\label{sec:pol}

In the present section we consider optical excitations between Landau levels of $\mathrm{MoS_2}$ and establish selection rules depending on the circular polarization of the radiation. To that effect, we assume that the $\mathrm{MoS_2}$ layer is exposed to circularly polarized light. We shall label $\kv_p$ the wave vector and $\hbar \omega$ the energy of the light field. $\kv_p$ is orthogonal to the plane of the material and way smaller than $1/\tilde{a} $, thus authorizing only vertical transitions. The polarization index is denoted as $\sigma$. For clockwise-polarized light $\sigma=+1$, otherwise $\sigma=-1$. We shall now determine interaction with light and the associated selection rules.

\subsection{General theory}
\label{sec:theopolar}
In order to take into account the coupling to the light field, one may again use the Landau--Peierls substitution with a new total potential $\Av_{\mathrm{tot}}=\Av+\Av_\mathrm{rad}(t)$, in which $\Av$ is the potential introduced earlier and 

\beq
\Av_\mathrm{rad} (t)=A
\begin{pmatrix}
\cos\left(\omega t \right)\\
\cos\left(\omega t-\sigma\frac{\pi}{2} \right)\\
0
\end{pmatrix},
\eeq
is the potential describing the light. The interaction between the light and electrons in the system is given by the Hamiltonian

\beq\label{eq:HamLM}
\Hhat_\mathrm{l}(t)=\frac{e}{\hbar}\nabla_{\kv}\Hhat_{\mathbf{B}}^{\xi, s}\cdot\Av(t)_\mathrm{rad}=\hat{W}_{\xi \sigma}\er^{-i\omega t}+\hat{W}_{\xi \sigma}^\dagger\er^{i\omega t},
\eeq 
which needs to be added to the Hamiltonian. Here,
\beq
\label{dipolemat}
\hat{W}_{\xi \sigma}=ev_F A \h_{\xi \sigma} \text{ with }
\h_{\xi \sigma}=\frac{1}{2}(\xi \sigma_x+\sigma i \sigma_y),
\eeq
where we have defined
\beq
\h_{\xi s =+1}=
\begin{pmatrix}
0 & 1\\
0 & 0
\end{pmatrix}
\text{ and }
\h_{\xi s =-1}=
\begin{pmatrix}
0 & 0\\
1 & 0
\end{pmatrix}.
\eeq

\begin{figure}[tb]
\centering
\includegraphics[width=5.5cm,angle=0]{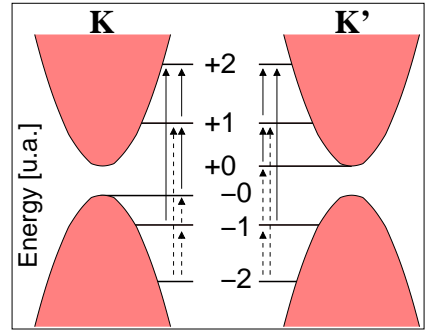}
\caption{\footnotesize{(Color online) Possible optical transitions between Landau levels
in the absence of spin-orbit coupling. Full arrows correspond to $\sigma=+1$ polarization whereas dashed arrows correspond to $\sigma=-1$ polarization.
}}
\label{fig:pol1}
\end{figure}

This may be be treated as a time-dependant perturbation, and transitions between initial states $\ket{i}$ and final states $\ket{f}$ are possible only if their respective energies are related to $ \omega$ by $E_f-E_i =\pm \hbar \omega$. The excitation term, which is what is interesting here, is proportional to $|\braket{f|\hat{W}_{\xi s}|i}|^2$ and thus to $|\braket{f|\h_{\xi \sigma}|i}|^2$. Hence, the transition rates are defined as
\beq
\mathcal{P}_{\lambda_i n_i,\lambda_f n_f}^{\xi, s,\sigma}=\left|\frac{\left(\psi^{\xi,s}_{\lambda_f n_f}\right)^\dagger \h_{\xi \sigma}{\psi^{\xi,s}_{\lambda_i n_i}}}{N_{\lambda_i n_i}^{\xi s} N_{\lambda_f n_f}^{\xi s}}\right|^2
\eeq
where we have explicitly taken into account the normalization (\ref{eq:normal})
of the vectors. The number $\mathcal{P}_{\lambda n_i,\lambda n_f}^{\xi, s, \sigma}$ is comprised between 0 and 1 which indicates the relative amount of electrons that will be excited for a given transition. It is thus a measure of the
strength of the associated absorption or emission peaks.

\subsection{Selection rules}

The above results can be used to determine which transitions are authorized for both polarizations in each valley. Considering the form of the vectors defined in  Eqs. (\ref{vectors1})-(\ref{vectors2}) and $h_{\pm 1}$, it is obvious that the only possible transitions are from states $\psi^{\xi, s}_{\lambda n_i}$ to $\psi^{\xi,s}_{\lambda n_f}$ such that $n_i$ and $n_f$ differ by exactly 1. Notice, however, that other transitions may occur if band corrections 
(such as trigonal warping) to the model are taken into account. We discuss these corrective terms in
more detail in Sec. \ref{sec:TW}.

All possible transitions as well as the corresponding amplitudes are given in Tab. \ref{tab:trans} and shown in Fig. \ref{fig:pol1}. It appears that the authorized transitions are the same in both valleys, with the exception of transitions implying the $\psi^{\xi,s}_{ 0}$ states. Hence, those are the transitions interesting for valley polarization. Other possible transitions which are activated at different energies are not considered in the following parts.

\begin{figure}[tb]
\centering
\includegraphics[height=6cm,angle=0]{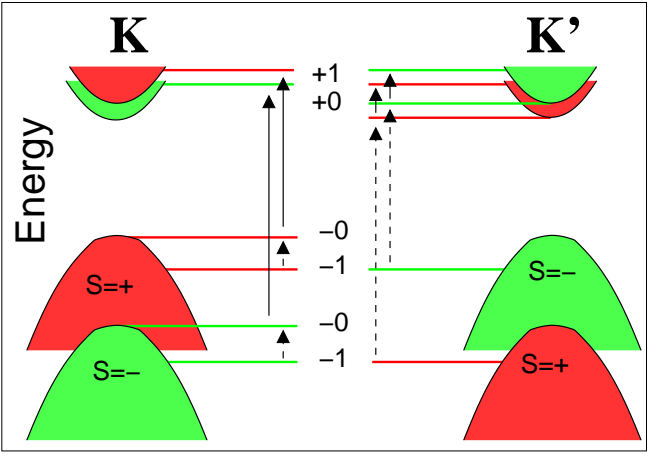}
\caption{\footnotesize{(Color online) Possible optical transitions in undoped \MoS{} involving $\psi^{\xi,s}_{\lambda 0}$ states between Landau levels in the presence of spin-orbit coupling. The splitting is reversed between the two valleys. $s=+1$ corresponds to the red bands while $s=-1$ is represented by the green bands. The arrows signification remains the same as in figure \ref{fig:pol1}. The figure takes into account the fact that $2\Delta \gg 2\Delta_\mathrm{so}^v \gg \delta_B^{\xi s},\Delta_\mathrm{so}^c$.}}
\label{fig:LLSO}
\end{figure}

If the Fermi level $\eps_F$ is comprised between $\eps_{-1}^{\xi s}$ and $\eps_{+1}^{\xi s}$, it is possible to polarize either valley using these transitions. To help characterize them, we define
\beqn\label{eq:apprGap}
\Delta_B^{\xi s}&=&\eps_{+1}^{\xi s}-\eps_{-0}^{\xi s} \approx 2\Delta_{\xi s}, \\ 
\delta_B^{\xi s}&=&\eps_{+1}^{\xi s}-\eps_{+0}^{\xi s} \approx \frac{\eps^2}{2\Delta_{\xi s}},
\eeqn
where we have used the fact that  $\eps \approx 30.5\, \sqrt{B(T)}$~meV~$\ll \Delta_{\xi s}$. $\Delta_B^{\xi s}$ is the energy associated with the $\psi^{\xi=+1,s}_{-0}\rightarrow\psi^{\xi=+1,s}_{+1}$ and $\psi^{\xi=-1,s}_{-1}\rightarrow \psi^{\xi=-1,s}_{+0}$ transitions, while $\delta_B^{\xi s}$ is associated with the $\psi^{\xi=+1,s}_{-1}\rightarrow\psi^{\xi=+1,s}_{-0}$ and $\psi^{\xi=-1}_{+0}\rightarrow\psi^{\xi=-1}_{+1}$ transitions.  The possible transitions involving the Landau
level $n=0$ are depicted in Fig. \ref{fig:LLSO}.

For the transitions discussed above, the relative amplitudes are readily calculated with
the help of the approximation (\ref{eq:apprGap}),
\beqn\label{eq:rays}
\nn
\mathcal{P}_{-0,+1}^{\xi=+1,s, \sigma=+1}&=&1-\frac{\eps^2}{4\Delta_{\xi s}^2}, \\
\nn
\mathcal{P}_{-1,-0}^{\xi=+1,s, \sigma=-1}&=&\frac{\eps^2}{2\Delta_{\xi s}^2}, \\
\nn
\mathcal{P}_{-1,+0}^{\xi=-1,s, \sigma=+1}&=&\frac{\eps^2}{4\Delta_{\xi s}^2},\\
\mathcal{P}_{+0,+1}^{\xi=-1,s, \sigma=-1}&=&1-\frac{\eps^2}{2\Delta_{\xi s}^2}.
\eeqn
As $\eps/\Delta_{\xi s} \approx 10^{-2}$, the magnitudes of the transitions involving $\sigma=-\xi$ polarizations are expected to be way less intense than $\sigma=\xi$ transitions.

\subsubsection{Transitions in undoped \textit{MoS}$_2$}

\begin{figure}[t]
\centering
\includegraphics[height=6cm,angle=0]{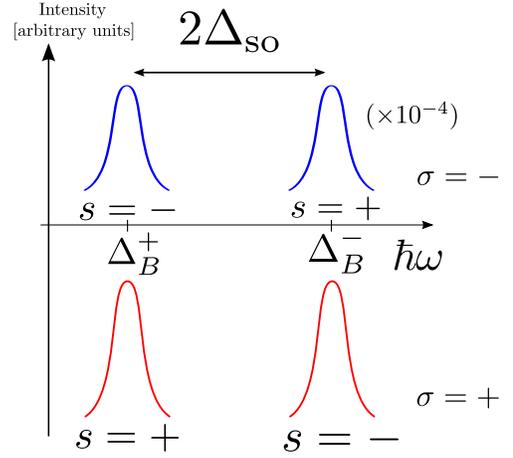}
\caption{\footnotesize{(Color online) Schematic plot of light absorption as a function of energy. The color of the peaks indicate the corresponding polarization (red for $\sigma=+$ and blue for $\sigma=-$). Each peak corresponds to excitations for a single valley and value of the spin. The peaks are separated by $\Delta_\mathrm{so} \approx 150$~meV. The peaks corresponding to $\sigma=-$ polarization are expected to be $10^4$ smaller than the peaks corresponding to $\sigma=+$ polarization.}}
\label{fig:peak}
\end{figure}

In the case of undoped \MoS, that is when the Fermi level is situated in the gap between the valance 
and the conduction band, the above analysis shows that it is possible to excite electrons in a single 
valley by the use of circularly polarized light, similarly to the case of \MoS~in the absence of a 
magnetic field.\cite{xiaoPrl} In contrast to the latter case, the magnetic field has two major 
consequences -- first, it defines well-separated energy levels that one may address in a resonant
manner; second, the absorption and emission peaks are proportional to the density of states, which
is strongly enhanced at resonance by the magnetic field because the density of states per Landau
level is given by the flux density $n_B\propto B$. As depicted in Fig. \ref{fig:LLSO}, light
with a polarization $\sigma=+$ is associated with the transition from $-0$ to $+1$ in the $\Kv$
valley, whereas light of polarization $\sigma=-$ couples the Landau levels $-1$ and $+0$ in
the $\Kv'$ valley. Furthermore, due to the spin-orbit gap, each transition is split into 
two rays $\Delta_B^{\xi s}$, such that one may furthermore identify each ray with a 
particular spin orientation of the involved electrons. This is depicted in Fig. \ref{fig:peak}.
The frequency of the rays is thus a direct measure of the spin-orbit gap in \MoS.
Notice finally that, as calculated in Eqs. (\ref{eq:rays}), the absorption peaks of light 
with polarization $\sigma=+$ (in the $\Kv$ valley) are much stronger than those for 
$\sigma=-$ (in the $\Kv'$ valley). This situation needs to be contrasted to the case of
silicene, where due to a strong spin-orbit gap the $n=0$ Landau levels are both situated at the 
bottom of the conduction band (for a particular spin orientation) such that circularly
polarized light excites electrons in both valleys, with roughly the same spectral 
weight.\cite{tabArx,tabArx2}

\subsubsection{Transitions in moderately doped MoS$_2$}

The transitions discussed in the previous paragraph are the only ones involving the $n=0$ level and
visible for undoped 
\MoS, i.e. when the Fermi level is situated in the gap between the valence and the conduction band. 
In the case of moderate doping, that is if the Fermi level $\eps_F$ is comprised between 
$\eps_{-1}^{\xi s}$ and $\eps_{+1}^{\xi s}$, other transitions involving $n=1$ and $0$ are 
possible. 
Indeed, using light of $\sigma=-$ polarization but of energy 
$\delta_B$ allows one to excite electrons in the valley $\Kv$, whereas a resonance at
$\Delta_B$ is still associated with a transition from $-1$ to $+0$ in the $\Kv'$ valley.
In the case of a polarization $\sigma=+$ the role of the valleys is exchanged.
Notice, however, that the resonances occur at extremely different energies since we have 
$\Delta_B^{\xi s} \approx 1.7$~eV, roughly independent of the magnetic field, whereas 
$\delta_B^{\xi s} \approx 0.27 \, B(T)$~meV is much smaller.

\section{Deviations from the magneto-optical selection rules due to band corrections}
\label{sec:TW}

In contrast to the preceding section, where massive Dirac fermions were considered, the band
structure of 2D \mos~reveals deviations from this ideal dispersion. The most prominent ones
are the electron-hole asymmetry, which yields a mass difference of roughly 20\% for electrons
and holes,\cite{zahArx} and trigonal warping.\cite{korArx} In the present section, we investigate
how these corrections affect the magneto-optical selection rules obtained above.

Similarly to monolayer graphene, trigonal warping arises from higher-order band corrections
beyond linear order in the off-diagonal terms and may be accounted for via the term\cite{goerRmp}
\begin{equation}\label{eq:tw}
\Hhat_\mathrm{3w} = 
\begin{pmatrix}
0 & \gamma \left (\xi q_x + i q_y\right )^2 \\
\gamma \left (\xi q_x - i q_y\right )^2 & 0
\end{pmatrix},
\end{equation}
whereas the electron-hole asymmetry is encoded in the corrective term
\begin{equation}
\Hhat_\mathrm{as} = 
\begin{pmatrix}
\alpha q^2 & 0 \\
0 & \beta q^2
\end{pmatrix}.
\end{equation}
The relevant parameters may be obtained from a fit to tight-binding or ab-initio calculations,
and one finds $\alpha = 1.72$~eV\AA$^2$, $\beta = -0.13$~eV\AA$^2$, 
and $\gamma = -1.02$~eV\AA$^2$.\cite{korArx}

\subsection{Modified Landau levels}

As in section \ref{sec:LL}, the modified Landau-level spectrum may be obtained with the 
help of the Landau--Peierls substitution. The term $\Hhat_\mathrm{as}$ remains diagonal in the 
basis of eigenstates of Eqs. (\ref{vectors1})-(\ref{vectors2}) and reads

\beq
\Hhat_{\mathbf{B},\mathrm{as}}^{\xi, s}=
\frac{1}{l_B^2}
\begin{pmatrix}
\alpha \left (2n+1\right ) & 0 \\
0 & \beta \left (2n+1\right )
\end{pmatrix}.
\eeq
Thus the eigenstates are of the same form as in Eqs. (\ref{vectors1})-(\ref{vectors2}),
with marginally different values of $ \alpha_{\lambda n}^{\xi s}$. However, the energies levels are slightly shifted 

\beqn
\eps_{\lambda n}^{\xi s}&=&\frac{\alpha+\beta}{l_B^2}\left(n+\frac{1}{2} \right )\\
\nn
&&+\lambda\sqrt{\left [\left (n+\frac{1}{2}\right )\frac{\alpha-\beta}{l_B^2}+\Delta_{\xi s} \right ]^2+n\eps^2 }\text{ for} \geqslant 1, \\
\eps_{n= 0}^{\xi s}&=& \frac{\alpha+\beta}{2l_B^2}-\xi\left( \frac{\alpha-\beta}{2l_B^2}+\Delta_{\xi s}\right ).
\eeqn

In contrast to the electron-hole asymmetry term, the trigonal wrapping 
term\cite{goerRmp,plochPrl}

\beqn
\Hhat_{\mathbf{B},\mathrm{3w}}^{\xi=+, s}&=&
-\frac{2\gamma}{l_B^2}
\begin{pmatrix}
0  & \left(a^\dagger \right)^2 \\
a^2 & 0
\end{pmatrix}
\\
\Hhat_{\mathbf{B},\mathrm{3w}}^{\xi=-, s}&=&
-\frac{2\gamma}{l_B^2}
\begin{pmatrix}
0  & a^2 \\
\left(a^\dagger \right)^2 & 0
\end{pmatrix}
\eeqn
is not diagonal in the basis (\ref{vectors1})-(\ref{vectors2}) and thus needs to be treated
perturbatively. Such a treatment shows that trigonal warping yields a second-order correction 
relative to the leading-order $(\hbar v_F/l_B)\sqrt{n}$ Landau-level 
behavior that arises only in second-order perturbation theory in $\gamma$.\cite{goerRmp,plochPrl}
However, the eigenstates are modified at first order and one finds that the original eigenstate
$\psi_{\lambda n}^{\xi, s}$ mixes with at most four states $\psi_{\lambda' n'}^{\xi, s}$ such that $ n'=n \pm 3$. The new eigenstate $\psit_{\lambda n}^{\xi, s}$ corresponding to the energy $\eps_{\lambda n}^{\xi s}$ is

\beq
\psit_{\lambda n}^{\xi, s}= \psi_{\lambda n}^{\xi, s}
+\sum_{\substack{\lambda'=\pm \lambda \\ n'=n\pm 3}}\frac{\left (\psi_{\lambda' n'}^{\xi, s}\right )^{\dagger}\Hhat_{\mathbf{B},\mathrm{3w}}^{\xi, s}\psi_{\lambda n}^{\xi, s}}{\eps_{\lambda' n'}^{\xi s} -\eps_{\lambda n}^{\xi  s}}  \psi_{\lambda' n'}^{\xi, s} 
\eeq

Since $|\eps_{-\lambda n'}^{\xi s} -\eps_{\lambda n}^{\xi s} |\approx \Delta$ while 
$|\eps_{\lambda n'}^{\xi s} -\eps_{\lambda n}^{\xi s}|\approx \delta_B^{\xi s} \ll \Delta$, 
the inter-band mixing with $\lambda'=-\lambda$ can be neglected in the sum. 
Evaluation of the matrix elements of $\Hhat_{\mathbf{B},\mathrm{3w}}^{\xi, s}$ yields

\beq\label{eq:statesPT}
\psit_{\lambda n}^{\xi, s}\simeq \psi_{\lambda n}^{\xi, s}
+\mu_{\lambda n}^{\xi s}\psi_{\lambda \left (n-3\right )}^{\xi, s}
+\nu_{\lambda n}^{\xi s}\psi_{\lambda \left (n+3\right )}^{\xi, s}
\eeq
with
\beqn
\mu_{\lambda n}^{+ s}&=& \delta_{n\geqslant 3}\frac{\gamma}{l_B^2}\frac{\left (\beta_{n-3}\right )^\dagger \alpha_{\lambda n}^{+ s}\sqrt{\left(n-1\right)\left(n-2\right)}}{N_{\lambda (n-3)}^{+ s}N_{\lambda n}^{+ s}\left(\eps_{\lambda (n-3)}^{+ s}-\eps_{\lambda n}^{+ s}\right )},\\
\mu_{\lambda n}^{- s}&=& \delta_{n\geqslant 3}\frac{\gamma}{l_B^2}\frac{\left (\alpha_{\lambda (n-3)}^{- s}\right )^\dagger \beta_{n}\sqrt{\left(n-1\right)\left(n-2\right)}}{N_{\lambda (n-3)}^{- s}N_{\lambda n}^{- s}\left(\eps_{\lambda (n-3)}^{- s}-\eps_{\lambda n}^{- s}\right )},\\
\nu_{\lambda n}^{+ s}&=&\frac{\gamma}{l_B^2}\frac{\left( \alpha_{\lambda\left (n+3\right )}^{+ s}\right)^{\dagger}  \beta_{n}\sqrt{\left(n+2\right)\left(n+1\right)}}{N_{\lambda n}^{+ s}N_{\lambda (n+3)}^{+ s}\left(\eps_{\lambda (n+3)}^{+ s}-\eps_{\lambda n}^{+ s}\right )},\\
\nu_{\lambda n}^{- s}&=&\frac{\gamma}{l_B^2}\frac{\left( \beta_{n+3}\right)^{\dagger}  \alpha_{\lambda n}^{- s}\sqrt{\left(n+2\right)\left(n+1\right)}}{N_{\lambda n}^{- s}N_{\lambda (n+3)}^{- s}\left(\eps_{\lambda (n+3)}^{- s}-\eps_{\lambda n}^{- s}\right )}.
\eeqn
where $\delta_{n\geqslant 3}$ symbolically indicates that $\mu_{\lambda n}^{\xi s}$ is
non-zero only for $n\geqslant 3$.
For these expressions to remain valid for the zero states one may define $\psi_{\lambda n}^{\xi=-\lambda, s}=0$ and $\alpha_{\lambda 0}^{\xi, s}=\beta_0=1$. Considering that $\gamma/l_B^2 \ll \alpha^{\lambda n}_{\xi s}$ or $\beta_n$ it is a good approximation to say that the norm $N_{\lambda n}^{\xi s}$ is unchanged for small values of $n$.

\subsection{Optical transitions}

With the help of the above-mentioned states, it is possible to examine the effect of the additional terms on the optical transitions. To that effect, one may use the same formalism as in Sec. \ref{sec:theopolar}. To take into account the addition of $\Hhat_\mathrm{as}$ and $\Hhat_\mathrm{3w}$, one has to change the $\hat{W}_{\xi \sigma}$ matrix of Eq. \eqref{dipolemat} into $\hat{W}_{\xi \sigma}^\mathrm{tot}=\hat{W}_{\xi \sigma}+\hat{W}^1_{\xi \sigma}$ with

\beqn\label{eq:LMPT}
\nn
\hat{W}_{+\xi }^1&=&
\frac{Ae}{l_B \hbar}
\begin{pmatrix}
\alpha i \sqrt{2} a^\dagger & -i \sqrt{2} \gamma (1-\xi)   a \\
i \sqrt{2} \gamma (1+\xi)  a^\dagger & \beta i \sqrt{2} a^\dagger
\end{pmatrix} ,\\
\hat{W}_{-\xi }^1&=&
\frac{Ae}{l_B \hbar}
\begin{pmatrix}
-\alpha i \sqrt{2} a & i \sqrt{2} \gamma (1+\xi)   a^\dagger \\
-i \sqrt{2} \gamma (1-\xi)   a & -\beta i \sqrt{2} a
\end{pmatrix}. 
\eeqn

Two types of corrections, to first order in $\gamma$ (and principally also in $\alpha$ and $\beta$),
need to be considered. First, the perturbed states (\ref{eq:statesPT}) allow for novel 
transitions when evaluated in the unperturbed coupling Hamiltonian (\ref{eq:HamLM}), 
due to the mixing between $\psit_{\lambda n}^{\xi, s}$ and $\psi_{\lambda (n\pm3)}^{\xi, s}$. 
In this case, the previous selection rules apply and thus, $n'=n\pm3 \pm 1$, i.e. $n'=n\pm 2$ or 
$n'=n\pm4$. Second,
the modified light-matter coupling $\hat{W}_{\xi\sigma}^1$ yields novel transitions when evaluated 
in the unperturbed states, such as for example the interband transition $n\rightarrow n$. These
transitions arise from the non-diagonal terms in Eq. (\ref{eq:LMPT}), whereas the diagonal terms yield
dipolar transitions $n\rightarrow n\pm 1$, as the ones discussed in Sec. \ref{sec:pol}.
Table \ref{tab:trans} can be used to determine the relative magnitude of the transitions. The value for the $\psit_{\lambda n}^{\xi, s}\rightarrow\psit_{\lambda' n'}^{\xi, s}$ transition relative amplitude with $ n'=n\pm2$ or 4 is the amplitude for $\psi_{\lambda (n\pm3)}^{\xi,s}\rightarrow\psi_{\lambda' n'}^{\xi, s}$ normalized with the adequate factor, either $ |\mu_{\lambda n}^{\xi s}/N_{\lambda n}^{\xi s} |^2$ or  $|\nu_{\lambda n}^{\xi s}/N_{\lambda n}^{\xi s}|^2$.

Other possible transitions involving both the perturbed states and the new light matrix elements are proportional to at most $l_B^{-4} \propto B^2$ and can thus be neglected to first order
in perturbation theory.

Henceforth, trigonal warping and electron-hole asymmetry induce additional transitions $\psi_{\lambda n}^{\xi, s}\rightarrow\psi_{\lambda' n'}^{\xi, s}$ with $n'=n$, $n'=n\pm2$ or $n'=n\pm4$. One may want to evaluate the relative intensity of corresponding absorption peaks, at least for small values of $n$. For transitions involving the $\hat{W}_{\xi \sigma}$ light matrix and perturbed $\psit_{\lambda n}^{\xi, s}$ states, the evaluation of $|\mu_{\lambda n}^{\xi s}/N_{\lambda n}^{\xi s} |^2$ shows that, for $B=10$~T, the $n'=n\pm3\pm1$ peaks should be about $3\,000$ times smaller than the regular peaks corresponding to $\psi_{\lambda (n\pm3)}^{\xi, s}\rightarrow\psi_{\lambda' n'}^{\xi, s}$ transitions. Similarly, the $n'=n$ peaks originating from the additional terms in the light coupling can be evaluated to be about $1\,000$ times smaller than the regular peaks.

\section{Conclusions}

In summary, we have used a two-band model that reduces to massive Dirac fermions
with a spin-valley dependent gap at low energies to investigate the magneto-optical properties
of \mos. Most saliently, the particular behavior of the $n=0$ Landau levels, which stick to the
top of the valence band and the bottom of the conduction band in the $\Kv$ and $\Kv'$ valleys,
respectively, allow for a selection of electrons in a particular valley via the circular 
polarization of the light field. Whereas the $-0\rightarrow +1$ transition (in the valley $\Kv$) 
is addressed by the polarization $\sigma =+$ the $-1\rightarrow +0$ transition (in the valley
$\Kv'$) couples only to light with a polarization $\sigma =-$. Moreover, because of the moderate
spin-orbit gap (mainly in the valence band), it is possible to address 
electrons with a particular spin 
orientation. Indeed, a resonant excitation of the above-mentioned Landau level transitions would allow
not only to excite electrons in a single valley (via the circular polarization of the light) but also
a single spin state in that valley because the resonance condition is spin-dependent. In light transmission
measurements of \mos{} flakes in a magnetic field, for example, one would therefore expect two absorption peaks 
for each polarization separated by the spin-orbit gap. This would allow for a direct 
spectroscopic measurement of the spin-orbit coupling in \mos{} in the vicinity of the $\Kv$
points. 

The analysis remains valid for other systems sharing the low-energy structure of \mos{}, 
as it might be the case for other group-VI dichalcogenides.\cite{xiaoPrl} Beyond the description
of low-energy electrons in \mos{} in terms of massive Dirac fermions, which yields the typical 
dipole-type magneto-optical selection rules $n\rightarrow n\pm 1$ (regardless of the bands involved),
we have shown that higher-order band corrections give rise to non-dipolar magneto-optical transitions. 
Whereas to first order in perturbation theory the Landau level spectrum is affected only by 
the particle-hole asymmetry, but not by trigonal warping, the latter induces novel transitions already
at first order. As such, we have identified the interband transition $n\rightarrow n$ as well as
$n\rightarrow n\pm 2$ and $n\rightarrow n\pm 4$. These transitions are expected to cause novel 
absorption peaks in light transmission experiments, albeit with a significantly lower spectral weight
as compared to the dipolar transitions.

\acknowledgments
We acknowledge fruitful discussions with Marek Potemski.

\end{document}